# Biological applications of the theory of birth-and-death processes

Artem S. Novozhilov, Georgy P. Karev, and Eugene V. Koonin*

National Center for Biotechnology Information, National Library of Medicine, National Institutes of Health, Bethesda, MD 20894

*To whom correspondence should be addressed. Email koonin@ncbi.nlm.nih.gov

**Abstract**

In this review, we discuss the applications of the theory of birth-and-death processes to problems in biology, primarily, those of evolutionary genomics. The mathematical principles of the theory of these processes are briefly described. Birth-and-death processes, with some straightforward additions such as innovation, are a simple, natural formal framework for modeling a vast variety of biological processes such as population dynamics, speciation, genome evolution, including growth of paralogous gene families and horizontal gene transfer, and somatic evolution of cancers. We further describe how empirical data, e.g., distributions of paralogous gene family size, can be used to choose the model that best reflects the actual course of evolution among different versions of birth-death-and-innovation models. It is concluded that birth-and-death processes, thanks to their mathematical transparency, flexibility and relevance to fundamental biological process, are going to be an indispensable mathematical tool for the burgeoning field of systems biology.

## Introduction

Mathematics has long been intertwined with the biological sciences. The importance of mathematical approaches in many areas of biology is obvious but it is less appreciated that biological questions have stimulated the emergence of a variety of new directions in mathematics. Among many others, the areas of mathematics fully or partially developed in response to demands of biology include branching processes, traveling wave solutions of reaction-diffusion systems, Turing bifurcation and diffusive instability, analysis of replicator equations, stochastic coalescent process, evolutionary game theory, and analysis of variance [1,2].

Another fruitful and diverse mathematical field inspired by biology is the theory of birth-and-death processes. This theory was developed in the beginning of the 20th century as a result of attempts to model growth of a population taking into account stochastic demographic factors. With time, the theory has been becoming increasingly sophisticated, spawning new branches of stochastic process analysis [3]. Importantly, however, the first and simplest birth-and-death processes considered by Yule [4], Feller [5], and Kendall [6] provide a natural and useful theoretical framework for several areas of modern biology, such as estimation of the age of alleles, reconstruction of phylogenies, and modeling various aspects of genome evolution.

A birth-and-death process is a stochastic process in which jumps from a particular state (number of individuals, cells, lineages etc) are only allowed to neighboring states (Fig. 1). A jump to the right, i.e, increase by one of the number of individual or similar quantities represents birth, whereas a jump to the left represents death (Fig. 1). This property considerably simplifies the mathematical analysis, but the process remains applicable to numerous real-world systems. Birth-and-death models allow one to address any questions formulated in terms of transition or state probabilities of the process, stationary distribution, mean, variance and distribution of times of the first entrance to a particular set of states, probabilities of extinction, the mean time of existence, etc. The results obtained with these models can be compared with empirical data allowing one to either reject some of the initial assumptions, or accept the model as a useful tool for analysis and prediction of properties of the real system.

Early models in population biology, including those based on the theory of birth-and-death processes, were largely deterministic. However, from the very beginning of population growth modeling, it had been clear that a more refined analysis must take into account the role of stochastic factors in the evolution of the population. An early recognition of this fact is evident in the 1874 study of extinction of human families by Watson and Galton [7] (reviewed, e.g., in [6]). The classic deterministic theory of population growth treats the size of a population as a continuous variable. This means that the state space of the process under consideration is continuous in the deterministic setting. By contrast, the state space of the corresponding stochastic process is discrete. In this regard, the stochastic models are more realistic than the deterministic ones because counts of individuals (genes, cells, number of species or gene families, etc.) are discrete by definition. Loosely speaking, each deterministic model can be viewed as an approximation of the corresponding stochastic model. This, however, does not mean that deterministic models always yield qualitatively valid solutions. A case in point is the phenomenon of persistence and its complement, extinction: there are situations when a deterministic model predicts that a population approaches a positive stationary level, whereas the corresponding stochastic model shows that extinction occurs with certainty. A classic example is a population model that consists of a simple branching process with the mean number of children equal to one. This process inevitably goes to extinction whereas the corresponding deterministic model describes a population with a constant size. Another well-known example is the logistic stochastic process considered below (see also, e.g., [8,9]).

Here we review the basic principles of the theory of birth-and-death processes and discuss examples of recent studies that involve, as the main or an auxiliary approach, analysis of a birth-and-death process. A simple introduction to the theory of birth-and-death processes is given in [9] and [8]. A more complete mathematical presentation can be found in [10-13] and in other mathematical texts on stochastic processes.

## Brief mathematical background

The general study of temporally continuous, stochastic models of population growth apparently started with the work of Feller [5]. The cardinal assumption was that the growth of a population can be represented by a Markov process, i.e., the state of the population at time $t$ can be described by the value of a random variable $X(t)$ with the property

$$\Pr\{X(t)=n \mid X(t_0)=m_0, X(\tau_1)=m_1,\ldots,X(\tau_k)=m_k\} = \Pr\{X(t)=n \mid X(t_0)=m_0\},$$

for all $\tau_i \le t_0$ and whenever $t_0 < t$. The nature of the variable $X(t)$ differs from model to model. In this review, we consider only models with continuous time although an analogous theory exists for stochastic birth-and-death processes with discrete time (e.g., [8]).

If we interpret $X(t)$ as a population size, then a birth-and-death process is a Markov process $\{X(t),\ t \ge 0\}$ such that, in an interval $(t, t+\Delta t)$, each individual in the population has the probability $\lambda_n \Delta t + o(\Delta t)$ of giving birth to a new individual (probability of transition from state $n$ to state $n+1$) and the probability $\mu_n \Delta t + o(\Delta t)$ of dying (probability of transition from state $n$ to state $n-1$). The parameters $\lambda_n$ and $\mu_n$ are called the birth rate and death rate, respectively ($n$ is the population size). For an intuitively plausible depiction of a birth-and-death process, it is useful to imagine a material particle, which moves from an integer to the neighboring integer, the path function $X(t)$ being the position of particle at time $t$ (Fig. 1).

The state probabilities $p_n(t) = \Pr\{X(t)=n\}$ of the process being in state $n$ at time $t$ satisfies the following system of differential equations, called Kolmogorov forward equations [9]:

$$\frac{dp_0(t)}{dt} = -\lambda_0 p_0(t) + \mu_1 p_1(t),$$
$$\frac{dp_n(t)}{dt} = \lambda_{n-1} p_{n-1}(t) - (\lambda_n + \mu_n) p_n(t) + \mu_{n+1} p_{n+1}(t), \quad n \geq 1 \tag{1}$$

Here we consider birth-and-death processes whose state space consists of non-negative integers $\{0,1,\ldots,N,\ldots\}$. Generally, there are two types of random processes: one in which there are no restrictions on the allowed set of states and the other in which there are restrictions in the sense that some states have special properties. For example, in a growing population, once the number $n$ of individuals is zero, the growth process stops (if there is no immigration). Thus the state $n = 0$ is a special state, i.e., once the process reaches this state, it is trapped forever. Such states are called absorbing states. Another special state is the so-called reflecting state. Once the process reaches a reflecting state, it must return to the previously occupied state. Having in mind biological applications of birth-and-death processes, we consider here random processes that have either one or two special (absorbing or reflecting) states.

Equations (1) have to be solved subject to an initial condition and some boundary conditions. It is sufficient to solve them for the initial condition $p_n(0) = \delta_{n,m}$, i.e., for the case when the process is initially in a definite state $m$. The state probabilities $p_n(t)$ give full information about the analyzed process but it is usually difficult to solve the system (1). The simplest case when the solution of system (1) is straightforward is a pure birth process or the Poisson process. In this case, we have $\lambda_n = \lambda,\ \mu_n = 0$, and the solution of (1) subject to the initial condition $p_0(0) = 1$ is the Poisson distribution

$$p_n(t) = \frac{(\lambda t)^n}{n!} e^{-\lambda t}$$

with parameter $\lambda t$.

It is well known that the distribution of the time intervals between any two successive jumps in any Markov process with continuous time and discrete space of states is exponential (e.g., [8]). More precisely, let $W_i$ be the instant of the $i$th jump of the birth-and-death process (1) and $T_i = W_{i+1} - W_i$ be the sojourn time; suppose that $X(W_i) = n$, then the process spends exponentially distributed time $T_i$ in the state $X(t) = n$ with the mean $E[T_i] = 1/(\lambda_n + \mu_n)$. When a jump occurs, it will be a birth with the probability $\lambda_n/(\lambda_n + \mu_n)$ or a death with the probability $\mu_n/(\lambda_n + \mu_n)$.

If birth and death rates of the system (1) are linear functions of $n$, then the so-called probability-generating functions technique can be applied for writing down the appropriate partial differential equations for the probability-generating function (e.g., [14]). Historically, this method had been the main tool for analyzing various birth-and-death processes [6,15]. It is straightforward to obtain results for a simple birth process ($\lambda_n = \lambda n, \mu_n = 0$ [9]), simple death process ($\lambda_n = 0, \mu_n = \mu n$ [9]), simple death-and-immigration process ($\lambda_n = \lambda, \mu_n = \mu n$ [16]), simple birth-and-death process ($\lambda_n = \lambda n, \mu_n = \mu n$ [9]), generalized simple birth-and-death process ($\lambda_n = \lambda(t)n, \mu_n = \mu(t)n$ [6]), and a simple birth-and-death process with immigration ($\lambda_n = \lambda n + \nu, \mu_n = \mu n$ [9,17]).

For example, for a simple birth process with the initial condition $p_m(0) = 1$, it can be shown that the state probabilities are

$$p_n(t) = \binom{n-1}{m-1} e^{-m\lambda t}(1 - e^{-\lambda t})^{n-m},\ n \geq m.$$

This stochastic process was first studied by Yule [4] in connection with the mathematical theory of evolution. The state of the process was thought of as a species within a genus, and the creation of a new species by mutation was conceived as being a random event with the probability proportional to the number of species. Yule used this process to explain the observed power law distribution of genera of plants having $n$ species.

For a simple birth-and-death process, one can find (e.g., [9])

$$p_0(t) = P_0,$$

$$p_n(t) = (1 - P_0)(1 - \lambda P_0 / \mu)(\lambda P_0 / \mu)^{n-1}, n \geq 1,$$

where $P_0 = \dfrac{\mu\left(e^{(\lambda-\mu)t} - 1\right)}{\lambda e^{(\lambda-\mu)t} - \mu}$,

under the initial condition $p_1(0) = 1$. For other possible initial conditions, the solution of (1) is more complicated but still can be obtained. Thanks to the exact nature of this solution, it has many applications in current research, e.g., phylogeny reconstruction phylogenies [18,19] and estimation of  the age of rare alleles [20].

As pointed out above, the method of probability-generating functions generally works when birth and death rates are linear functions of $n$. Karlin and McGregor [17,21] showed that the solution of (1) could be obtained with the help of a sequence of orthogonal polynomials, which are closely related to the birth-and-death process. The general linear case $\lambda_n = \lambda n + \nu$, $\mu_n = \mu n + \rho$ was solved in [17] for $\rho = 0$. The case of $\rho \neq 0$ was analyzed in [22]. The asymptotically symmetric quadratic case $\lambda_n = (N - n)(n + a)$, $\mu_n = n(n + b)$ first appeared in applications concerned with genetic models [23]. Some other special cases also have been described [24], and more on recent advances in the study of birth-and-death processes and associated polynomials can be found in [25]. The problem with exact solutions of system (1) is that, in many cases, the expressions for the state probabilities, although explicit, are intractable for analysis and include special polynomials. In such cases, it may be sensible to solve more modest problems concerning the birth-and-death process under consideration, without the knowledge of the time-dependent behavior of state probabilities $p_n(t)$.

It is usually not difficult to write down the differential equations for the first few moments of $X(t)$ (e.g., [14]). For example, for a simple birth process, the equation for the mean $E[X(t)]$ is

$$\frac{dE[X(t)]}{dt} = \lambda E[X(t)]$$

with the solution $E[X(t)] = e^{\lambda t}$ for the case $p_1(0) = 1$. It should be noticed that, in this case, the mean growth of the process follows the same exponential law as the one that appears in the simplest deterministic model of population growth, namely, $dN(t)/dt = \lambda N(t)$, where $N(t)$ is the population size and $\lambda$ is a Malthusian parameter. Sometimes, this fact is used as justification for the assertion that the deterministic theory is simply an account of the expectation behavior of the random variables, which occur in the stochastic formulation. This is not generally true as first pointed out by Feller [5]. For example, this is not the case for the logistic stochastic process considered below.

The method for writing down equations for the moments of $X(t)$ works only when $\lambda_n$ and $\mu_n$ are linear functions of $n$. If $\lambda_n$ and $\mu_n$ include terms with $n$ of degree higher than 1, the equation for the mean involves, generally, the second moment, the equation for the second moment involves the third moment and so on. This hierarchy of equations can only be solved approximately, by making any of a variety of approximations. Recently, the moment closure approximation has become a popular technique for obtaining such solutions [26]. Applications of this method to the logistic stochastic process can be found in [27,28].

Some other tractable quantities give insight into the evolution of a birth-and-death process, in particular:

i) the steady-state probabilities $p_n^* = p_n(\infty)$ describing the process when it is in a dynamical equilibrium;
ii) the probability that a given state is ever reached;
iii) the "first passage time", i.e., the time for the process to reach a given state for the first time, its probability density function, and its moments.

A full account of the possible expressions for these quantities is given in [14]. These formulas are exact but in many cases computationally intensive; however, for many of these expressions, useful approximations can be found, and the availability of exact formulas allows one to test these approximations.

## Models of population evolution based on the birth-and-death process
### The Moran model

One of the main goals of theoretical population genetics is to study changes in the genetic makeup of the population under various evolutionary forces, such as selection, mutation, migration and random drift. Theoretical studies rely on deterministic and stochastic mathematical models which capture the essential features of populations. Deterministic models are applicable when the size of the population is very large; and when the finite population size cannot be ignored, stochastic models should be used. Furthermore, certain problems can be solved only in stochastic settings.

Let us consider a population of haploid individuals of a fixed constant size $N$ (this is not a very restrictive assumption (e.g., [29]). Let us further consider one locus that can have two types of alleles, $A$ and $a$. The individuals with different alleles may differ in fitness, i. e., some individuals have a selective advantage over others. A few of the many problems intrinsic to this model that can be of a particular interest are the following. Because of random drift (stochastic nature of births and deaths), one of alleles eventually goes extinct; if there are no mutations, the population becomes homogeneous, and the interesting quantity is the speed with which it approaches homozygosity. Another situation is appearance of a unique mutant (e.g., $A$); in this case, the probability of fixation of $A$ and the mean time of fixation are of particular importance. If we assume that mutations can go in either direction, then the population will be heterozygous forever, and the variable of interest is the stationary distribution. To study all these problems quantitatively, a mathematical model is required.

Two such models have been the basis of most work in population genetics: the Wright-Fisher model [30,31] and the Moran model [29]. The Wright-Fisher model describes populations with discrete, seasonal reproduction and non-overlapping generations, whereas the Moran model is most applicable to populations with continuous reproduction. One of the possible formulations of the Moran model represents a continuous time birth-and-death process with nonlinear rates and finite state space. The Moran model is important for two reasons. First, in contrast to the Wright-Fisher model, it applies to organisms with overlapping generations. Second, many results that can be obtained only approximately under the Wright-Fisher model can be derived exactly using the Moran model. As Moran pointed out, “this model is simpler because from any state one can move only to the neighboring ones, which simplifies the theory” [31].

In order to analyze the model, we need to define all the birth and death rates. Let there be $n$ copies of allele $A$ and $N-n$ copies of allele $a$; each individual lives for an exponentially distributed amount of time with a mean of 1 and then is “replaced”; the replacement for individual $x$ is chosen at random from all individuals in the population including $x$ itself; the new individual replaces the old one at $x$. It is assumed that individuals carrying $A$ have the selection coefficient $s$. The transition rates for the Moran model can be written as

$$\lambda_n = (1+s)\frac{N-n}{N}p_n, \quad \mu_n = \frac{n}{N}(1-p_n), \tag{2}$$

where $p_n$ is the probability that the choice results in an $A$ if there are $n$ copies of this allele in the population. Assuming that there are no mutations, $p_n = n/N$. If we assume that the mutation rate from $A$ to $a$ is $v$, and from $a$ to $A$ is $u$, then

$$p_n = \frac{n}{N}(1-v) + \frac{N-n}{N}u .$$

If $v \neq 0$ and $u \neq 0$, we deal with a birth-and-death process with reflecting boundaries. The presence of reflecting boundaries means that there exists a stationary distribution $p^*$ which is easy to calculate numerically noting that, at equilibrium, $p^*_n \mu_n = p^*_{n-1}\lambda_{n-1}$ must be satisfied. Under particular conditions (when $N$ is large), a good approximation for the stationary distribution can be found (e.g., [10]).

If we assume that there are no mutations, then $\lambda_0 = \mu_N = 0$, and we have a birth-and-death process with absorbing boundaries. In this case, the fate of a unique mutant appearing in the population can be studied. One of the main questions in this situation is the probability of fixation, i.e., the probability that a new mutant penetrates the entire population. In mathematical terms, this can be expressed as the probability of reaching the absorbing state $N$ before reaching the absorbing state 0. The probability that the system ends up in the state $N$ (the probability of fixation) if initially there is only one $A$ is (e.g., [14])

$$P_{fix} = \frac{1}{1 + \sum_{i=1}^{N-1} \prod_{n=1}^{i} \frac{\mu_n}{\lambda_n}} .$$

Using (2) with $u = v = 0$, it is readily evaluated to

$$P_{fix} = \rho(s) = \frac{1-(1+s)^{-1}}{1-(1+s)^{-N}} \approx \frac{1-e^{-s}}{1-e^{-sN}} , \qquad (3)$$

where the second formula in (3) is the classical formula for the probability of fixation of a selective allele obtained by Kimura from the diffusion approximation of the Wright-Fisher model [32]. The more common form of the last formula has the factor $2s$ rather than $s$. The difference comes from different samplings in these two models and, accordingly, different rates of random genetic drift [33]. If $s = 0$, one can find that $P_{fix} = 1/N$, the probability of fixation of a neutral mutant.

**Logistic growth**

Since the very beginning of the study of birth-and-death processes, the logistic process had been considered [5]. The deterministic version of the logistic model was originally introduced by [34]. This model accounts for the density dependence in the growth of a single population. It is based on the hypothesis that the net birth rate per individual (i.e., the difference between the birth rate and the death rate) is a linearly decreasing function of the population size. This implies that the net population birth rate is a quadratic function of the population size. The model is closed in the sense that no immigration or emigration is allowed. Mathematically, the deterministic logistic model leads to a nonlinear differential equation

$$\frac{dN(t)}{dt} = rN(t)\left(1 - \frac{N(t)}{K}\right), \quad N(0) = N_0 \tag{4}$$

that can be solved explicitly. Here $N(t)$ is the density of the population at moment $t$, $r > 0$ is the intrinsic growth rate, and $K > 0$ is the carrying capacity. All solutions of equation (4) monotonically lead to the asymptotically stable equilibrium $N^* = N(\infty) = K$.

Various stochastic formulations of the logistic model have been developed and analyzed (see [8,35]). Apparently, the first stochastic version of the logistic model was formulated by Feller [5] as a finite-state birth-and-death process. Feller derived the Kolmogorov forward equations for the state probabilities. One of his results is that the solution of the deterministic model does not agree with the expectation of the solution of the stochastic model when both are studied as functions of time. However, the difference can be shown to be asymptotically small as the maximum population size increases.

The logistic stochastic process is important for several reasons. It is well appreciated that the genetic makeup of a population strongly depends on the population structure while most of the population evolution models (like the Moran model considered above) assume a constant population size. So, for more realistic modeling of population evolution, stochastic models with changing population size are required. Density-dependant effects influence the size of the population, preventing indefinite growth, and the logistic model is the simplest stochastic model with changing population size and density-dependant mechanisms that affect this size.

In the logistic model, the state zero is usually an absorbing state such that eventual absorption at the origin is certain (contrary to the solution of the deterministic model), and all states except the origin are transient (the state is called transient if the process visits this state only finitely many times). Two qualitatively different behaviors are possible at any given time: the process either goes extinct after having reached the absorbing state at the origin or remains in the set of transient states. Thus, the immediate two issues to address are the calculation of the mean time to extinction and the possible behavior of the system prior to extinction. The mean time to extinction can be calculated using the known formulas. For example, Goel and Richter-Dyn use a stochastic version of a special case of the Verhulst model to examine the extinction of a colonizing species[14]. However, the time to extinction may not have a known distribution, because of which characterizing the system by the mean time to extinction can be misleading.

The behavior of the process prior to extinction can be productively explored within the framework of the so-called quasi-stationary distributions [36]. The quasi-stationary distribution cannot be found analytically but there are effective numerical methods for determining such distributions [35]. The simplest way to obtain an approximation of the quasi-stationary distribution is to restrict consideration to transient states (making the state space strictly positive). By excluding zero from the state space, one can establish a related process without an absorbing state. This method has been applied in several mathematical models [15,37] and is valid when the time to extinction is reasonably large [35]. More recent results on stochastic logistic process can be found in [27,28,38-40].

# Applications

### Rates and patterns of gene duplications and evolution of multigene families

Gene duplication is one of the principal mechanisms of genome evolution [41,42]. In particular, in multicellular eukaryotes, gene duplications gave rise to numerous multigene families, and the mode of evolution of genes within multigene families has been the subject of many theoretical and empirical studies. These studies reveal a high degree of within-species homogeneity among duplicated sequences. It has been proposed that a "correction" mechanism must have operated to spread mutations among the paralogous genes in a multigene family, a phenomenon dubbed concerted evolution [43-45]. Concerted evolution essentially means that members of a multigene family do not evolve independently of each other. A classic example of this type of gene family is the cluster of rRNA genes in which all several hundred paralogous genes have (nearly) identical sequences, even in nontranscribed spacer regions [46]. Thus, the genes of this cluster in humans are more similar to one another than to most of the rRNA genes of chimpanzees. This high degree of sequence homogeneity within species is believed to be achieved by gene conversion, i.e., frequent interlocus recombination.

However, not all patterns of variation between genes in multigene families can be explained by concerted evolution [44,47]. The best characterized cases in point are the the multigene families of major histocompatibility complex (MHC) genes and immunoglobulin genes (Ig). The members of these families from the same species are not necessarily more closely related to one another than to the genes from different species. To characterize the mode of evolution of these systems, Nei and coworkers conducted phylogenetic analyses of several multigene families of the MHC and Ig systems [47]. The results show that the evolutionary pattern of these families is quite different from that of concerted evolution. To explain this distinct pattern, a birth-and-death model of evolution has been proposed [47,48]. Under this model, paralogous genes are produced by various mechanisms, including tandem and block duplication, and some of the duplicates diverge functionally whereas others become pseudogenes and eventually deteriorate beyond recognition. The end result of this mode of evolution is a multigene family consisting of a mixture of divergent groups of genes, with highly similar sequences within groups plus a substantial number of pseudogenes. The main difference between the birth-and death evolution process and concerted evolution is that, under the former, genes in multigene families evolve independently of each other. The results of phylogenetic studies on several other gene families, including ubiquitins, histones and various receptors, are consistent with the predictions of the birth-and-death evolution model [49-53].

The birth-and-death evolution model of multigene family evolution, although originally presented at the verbal level, allows the use of the classical theory of birth-and-death processes to estimate the rates of gene duplication and loss for groups with a well-established phylogeny; in particular, such an analysis has been performed for vertebrate evolution over the last 500 million years [54]. This problem is closely related to the problem of phylogeny reconstruction. It has been shown that it is possible to estimate the birth and death rates of lineages for reconstructed phylogenies although they contain no explicit information about extinct lineages [18,55]. A null model in phylogenetics is usually one in which, over a short time span, each extant lineage has the same probability of dividing into two or dying as any other lineage. Assuming that the birth and death rates are constant, this model is a continuous-time, linear birth-and-death process with transition rates $\lambda_n = \lambda n$ and $\mu_n = \mu n$. The birth-and-death model relates the number of extant lineages $N_T$ and the expected number of lineages $N_t$ at time *t,* which survive to or at least have one descendant at time $T$, by the formula [56]:

$$\frac{N_t}{N_T} = \frac{\lambda - \mu}{\lambda \exp\{(\lambda - \mu)(T - t)\} - \mu}.$$

If phylogeny can be reconstructed from the available data, it is possible to produce a lineage-through-time plot (number of extant lineages versus time). Fitting the model to the plot (e.g., by the least square method) allows estimating the rates of lineage birth and death.

Cotton and Page applied this model to estimate rates of gene duplication and loss [54]. They showed that constant rates of gene duplication and loss fit the pattern of recent gene family evolution reasonably well, implying that, contrary to several recent suggestions [57-59], there is no evidence of a recent increase in duplication rate. The appearance of such an increase is likely to be an artifact caused by the fact that a greater number of recent duplications have survived in extant genomes compared to older duplications [18]. By contrast, the second, ancient peak of duplications at ~500 million years ago, originally detected by Gu and coworkers [58], could not be accounted for by the constant-duplication-rate model and is likely to reflect a whole-genome (2R) or, at least, very large scale duplication in the early evolution of vertebrates [54]. The estimates of both duplication and loss rates obtained by Cotton and Page (duplication rate of ~0.00115 $\text{Myr}^{-1}\,\text{lineage}^{-1}$ and a loss rate of ~0.00740 $\text{Myr}^{-1}\,\text{lineage}^{-1}$) are substantially (about an order of magnitude) lower than previous estimates [42].

**Origin of power law distributions of genome-related quantities**

A broad variety of phenomena in physics, biology, and the social sphere is described by power law distributions[60-62]. In the field of genomics, the power laws have been observed in the distribution of the number of transcripts per gene, the number of interactions per protein, the number of genes in coexpressed gene sets, the number of genes or pseudogenes in paralogous families, the number of connections per node in metabolic networks, and other quantities that can be obtained by genome analysis [61,63-67]. These distributions are described by the formula $P(i) = ci^{-\gamma}$, where $P(i)$ is the frequency of nodes with exactly $i$ connections or sets with exactly $i$ members, $\gamma$ is a parameter which typically assumes values between 1 and 3, and $c$ is a normalization constant. More recently, it has been shown that the distributions of several genome-related quantities are best described by the generalized Pareto function $P(i) = c(i + a)^{-\gamma}$, where $\gamma > 0$, $a$ are parameters [68-70]. At large $i$ ($i >> a$), this distribution is indistinguishable from a power law, but at small $i$, it deviates substantially, with the magnitude of the deviation depending on $a$.

As already pointed out in the preceding section, in a somewhat different context, a birth-and-death process is a natural mathematical framework for modeling evolution of gene families, with .duplication constituting a gene birth and gene loss treated as a death event. The birth-and-death approach has been applied to modeling the evolution of paralogous genome family sizes [65,70,71], the distribution of folds and families in the protein universe [72], and protein-protein interaction networks [73].

For the analysis of the evolution of gene family sizes, which we consider here in somewhat greater detail, a third elementary process, innovation, via horizontal gene transfer or emergence of genes from non-coding sequences, has been incorporated into the model. Innovation is analogous to birth except that an innovation event specifically adds a member to class 1 (Fig. 1) which corresponds to the acquisition of a new gene by the analyzed genome. Accordingly, this type of evolutionary models has been dubbed birth-death-and-innovation models (BDIM) [70]. Following [70], let us suppose that we deal with a general birth and death process with state probabilities given by system (1). Assume that $\lambda_N = 0$ where $N$ is the maximal possible mutligene family size. This means that the process under consideration is a general birth-and-death process with finite state space and reflecting boundaries (i.e., $\lambda_0 \neq 0$, $\mu_N \neq 0$). Consequently, there is a stationary equilibrium distribution $p^*$. Let us define a function $\chi(n) = \lambda_{n-1} / \mu_n$. The asymptotic of the stationary distribution is completely defined by the asymptotic behavior of $\chi(n)$.

Let us suppose that, for large $n$, the following expansion is valid:

$$\chi(n) = \frac{\lambda_{n-1}}{\mu_n} = n^s \theta\left[1 - \gamma / n + o(1/n)\right],$$

where $s$ and $\gamma$ are real numbers and $\theta$ is positive. The following main result gives classification of possible stationary distributions of the birth-and-death model ([70], theorem 1):

*(i)* *If* $s \neq 0$ *then* $p_n^* \sim \Gamma(n)^s \theta^n n^{-\gamma}$, *where* $\Gamma(n)$ *is the gamma-function;*

*(ii)* *If* $s = 0$ *and* $\theta \neq 1$, *then* $p_n^* \sim \theta^n n^{-\gamma}$;

*(iii)* *If* $s = 0$, $\theta = 1$ *and* $\gamma \neq 0$, *then* $p_n^* \sim n^{-\gamma}$;

*(iv)* *If* $s = 0$, $\theta = 1$ *and* $\gamma = 0$, *then* $p_n^* \sim 1$.

This theorem implies that, if the birth and death rates of the process are balanced, in such a way that expansion of $\chi(n)$ has the form $\chi(n) = 1 - \gamma / n + o(1/n)$, then the resulting stationary distribution asymptotically tends to a power law. Analysis of the fit of different versions of the BDIM to the empirical distributions of gene family size in a variety of genomes showed that the simplest form of BDIM that gave a good fit to the empirical data was the second-order balanced linear model [$\lambda_n = \lambda(n + a)$, $\mu_n = \mu(n + b)$], whereas the simple model $\lambda_n = \lambda n$, $\mu_n = \mu n$ could be confidently rejected ([70] and Fig. 2). This result implied that the evolution of individual genes within a family of paralogs was not entirely independent, i.e., there was at least a weak dependence of gene birth and death rates on family size (or an “interaction” between paralogs, by analogy to chemical kinetics).

In a further development of this analysis, the characteristics of evolution of the system, such as the probability of formation of a family of the given size before extinction and the mean times of formation and extinction of a family of a given size, were examined by analyzing the stochastic version of BDIM [74,75]. Given the published estimates of the rates of gene duplication and loss [42], it has been found that the linear model, which gives a good approximation of the stationary distributions of family sizes for different genomes, predicts completely unrealistic mean times for reaching the observed sizes of the largest gene families. In computer simulations with a large ensemble of genes, even the minimum time required for the formation of the largest family has been shown to be unrealistically long. Thus, the linear BDIM is incompatible with the estimates of the rate of genome size growth derived from the empirical data (the use of the lower estimates of duplication rates reported by Cotton and Page [54] would only exacerbate the problem). By contrast, models with the degree between 2 and 3 [e.g., for the model of degree 2, or the quadratic model, $\lambda_n = \lambda(n + a_1)(n + a_2)$, $\mu_n = \mu(n + b_1)(n + b_2)$ ], which involve stronger dependences between the evolutionary rates of paralogs, resulting in a self-accelerated evolutionary regime, could account for faster evolution, although the mean times of formation of the largest gene families are still too long to fit the actual time scale of evolution. However, it has been shown that the variance of the time required for the formation of the largest families is extremely large (coefficient of variation $>> 1$), which means that some families would grow much faster than the mean rate. Thus, the minimal time required for family formation is more relevant for a realistic representation of genome evolution than the mean time. Monte Carlo simulations of family growth from an ensemble of simultaneously evolving singletons show that the time elapsed before the formation of the largest family was much shorter than the estimated mean time and approached realistic values [74,75] (Fig. 3). The biological interpretation of the dependence between evolutionary rates of paralogs that is intrinsic to the higher-degree BDIM, is not immediately obvious. However, it is tempting to propose that such dependence reflects the fact that evolution of paralogous families, particularly, large ones is an adaptive process governed by positive Darwinian selection. Birth-and-death models do not include selection explicitly but the requirement for higher-degree BDIM to present a realistic description of genome evolution may imply the decisive role of selection in this process.

Almost simultaneously, another model of gene family evolution generating the observed power law distributions has been proposed [76,77]. Here, the process under consideration is a simple linear birth-and-death process with immigration. With a given intensity $\rho$, there is an incoming Poisson process which corresponds to appearance of new gene families, supposedly, via duplication of a pre-existing gene followed by radical modification. It has been shown that, asymptotically, power law distribution for the state probabilities is possible if $\lambda > \mu$ [76,77]. In the model of Karev and coworkers [70,74,75], the condition is the opposite, i. e., a balanced BDIM with a power law asymptotic can be obtained only if $\lambda < \mu$, which is in full agreement with the latest estimates of duplication and loss rates [54].

**Modeling horizontal gene transfer**

Comparison of the numerous sequenced genomes from all three kingdoms has shown beyond reasonable doubt that horizontal gene transfer (HGT) occurred on numerous occasions, at least, in the evolution of prokaryotes and unicellular eukaryotes [78-82]. However, rigorous quantitative evaluation of the amount of HGT is extremely hard which leads to vast differences in opinion on the role of HGT in evolution, from the concept of pervasive HGT fully shaping prokaryotic genomes [82] to a much more reserved view of HGT as a relatively minor evolutionary factor [83-85]. The first evolutionary-theoretical analysis of horizontal gene transfer in microbial populations has been reported by Berg and Kurland who applied the Moran model for the analysis of genetic drift in a haploid population [86]. A more general model of horizontal gene transfer, which includes both intra- and inter-population gene fluxes, has been recently formulated and analyzed [87]. In particular, this work involved an attempt to identify the conditions under which a horizontally transferred sequence can be fixed or at least penetrate a significant fraction of the population.

The model includes five parameters: inactivating mutation rate $u$ of the novel sequence, selection coefficient $s$, invasion rate $\gamma$ (the rate with which the individuals of the population acquire a new gene from other populations), within-population horizontal transmission (infection) rate $\theta$, and population size $N$. The state of the process $n$ represents the number of individuals carrying the novel sequence. The rates are

$$\begin{aligned} \lambda_n &= \left[(1+s)(1-u)n + \gamma N\right]\frac{N-n}{N+1} + \theta\frac{n(N-n)}{N} \\ \mu_n &= \left[N-n+u(1+s)n\right]\frac{n}{N+1} \end{aligned} \quad (5)$$

According to the rate equations (5), we deal with a birth-and-death process with a finite state space and reflecting boundaries (i.e., $\lambda_0 \neq 0, \mu_N \neq 0$). This means that there is a stationary distribution for which a good approximation can be found. This stationary distribution can be used to estimate the mean population penetration of the novel sequence depending on the parameter values (Fig. 4). Figure 4 shows that, if the rate of invasion $\gamma N$ is substantially lower than the rate of inactivating mutation $uN$, significant penetration (on average) can be reached only with high positive values of $(s+\theta)N$, i. e., when the new gene is rapidly spread within the population and/or confers a selective advantage onto the recipient.

Berg and Kurland [86]considered the model with rates (5) and $\gamma = 0$ and came to the conclusion that horizontally acquired new genes can be fixed in a population only when these sequences confer a substantial selective advantage onto the recipient and therefore are subject to strong positive selection. If there is no invasion in the model, the new gene is doomed to extinct. But a more detailed analysis of the mean time to extinction and quasi-stationary distributions shows that a new gene can penetrate a significant part of the recipient population on its way to extinction [87]. Taking into account the processes of within-population transmission (infection) and invasion leads to conclusions that are substantially different from those of Berg and Kurland: if the rates of these processes are non-negligible, horizontally transferred sequences do get fixed or at least persist in a significant part of the recipient population for a long time, even if they are neutral or slightly deleterious. In other words, complete analysis of the state space given by the equations (5) shows that fixation or persistence of horizontally transferred genes in a population can be achieved via different routes: it can happen either due to a high rate of invasion or a high rate of infection, or substantial selective value. Generally, the modeling results are compatible with the notion that HGT, while certainly not completely promiscuous or uniformly rampant, could be a pivotal force in evolution, at least in the prokaryotic world.

**Somatic evolution of cancer cells**

Recently, the Moran model has been applied to model carcinogenesis [88-92]. The main assumption in these studies is that cancer is initiated in tissue compartments where only a relatively small number of cells are at risk of mutating into a transformed state escaping homeostatic regulation. In this case, the evolutionary dynamics can be approximated by a low-dimensional stochastic process with a linear Kolmogorov forward equation that can be solved analytically. Most of the time, the cell population is homogeneous with respect to the relevant mutations.

First, assume that there are two types of cells, A and B, and let *a* and *b* be their numbers. Cells can reproduce, mutate or die. Mutation is allowed from A to B with rate *u*. The total number of cells is kept constant (*N*), *N*=*a*+*b*. Cells can have different fitness. Under these assumptions, evolution of cells can be described with the Moran process; the transition matrix is given by

$$P_{ij}=\begin{cases}\dfrac{u(N-i)+ri}{N-i(1-r)}\dfrac{N-i}{N}, j=i+1,\\ \dfrac{(1-u)(N-i)}{N-i(1-r)}\dfrac{i}{N}, j=i-1,\\ 1-P_{i,i+1}-P_{i,i-1}, j=i,\\ 0, otherwise.\end{cases}$$

The corresponding Markov process is a biased random walk with one absorbing state, *b*=*N*. If $u \neq 0$, then, sooner or later, the system finds itself in this absorbing state. Under conditions given in [88], the probability of finding the system with *b* cells, 1<*b*<*N*, is of the order of *u* <<1/*N*. The system spends most of the time in the homogeneous states *b*=0 and *b*=*N*. If we denote the state with all *a* as *A* and the state with all *b* as *B*, then, approximately, we have the two-state chain

$A \xrightarrow{uN\rho(s)} B$,

where $\rho(s)$, is the probability of fixation (3); $uN\rho(s)$ is the corresponding rate of jump from one (almost) homogeneous state to another. This stochastic process is described by the following Kolmogorov forward equations:

$$\mathrm{d}P_A/\mathrm{d}t=-uN\rho,\ \mathrm{d}P_B/\mathrm{d}t=uN\rho,\ A(0)=1,\ B(0)=0.$$

The exact conditions when this approximation is valid have been determined [88].

The inactivation of both alleles of a tumor suppressor gene (TSG) implies a net reproductive advantage for the affected cell and might lead to clonal expansion. The inactivation of the first allele of the TSG may be neutral or may lead to a selective advantage or disadvantage. Thus, the process of inactivation of a TSG can be modeled by the system with three types of cells *a*, *b* and *c*. The mutational network can be depicted as:

$a_{(1)} \xrightarrow{u_1} b_{(s_1)} \xrightarrow{u_2} c_{(s_2)}$,

where $s_i, i=1,2$ are selection coefficients. Again, the total number of cells is kept constant, $N=a+b+c$. In the context of this model, one may be interested in estimating the mean time to fixation for the third type of cells (e.g., those with both TSG alleles inactivated).

To approximate the stochastic process under consideration, one should take into account the possibility of jumping from the state *A* to the state *C* (a stochastic tunnel). It can happen if a mutation from *a* to *b* gives birth to a lineage of cells of type *b* that does not go to fixation (these cell may go extinct due to random drift combined with selection pressure, especially if $s_1 < 1$), at the same time giving birth to a lineage of cells of type *c.*

The same kind of model was used to answer the question whether genetic instability (an elevated mutation rate) is an early (i.e., a potential cause) or late (i.e., probable side-effect) event in tumorigenesis. The path to cancer cells can be described by the following mutational network:

$$\begin{array}{ccccc} X_0 & \to & X_1 & \to & X_2 \\ \downarrow & & \downarrow & & \downarrow \\ Y_0 & \to & Y_1 & \to & Y_2 \end{array}$$

Here $X_0$ are cells in which both copies of a TSG are functional, $X_1$ are cells with one functional and one inactivated copy of TSG, $X_2$ are cells with both copies of TSG inactivated, and $Y_i$ are cells with zero, one or two inactivated TSG alleles and genetic instability that can be caused by mutation in one of $n_c$ genes. The possibility of stochastic tunnels complicates the analysis but all possible cases can be classified [88]. Mathematically, the question of whether genetic instability is an early or a late event is answered in terms of relative values of $X_2$ and $Y_2$. In the work of Nowak and coworkers, the threshold value of genes that can induce genetic instability is found in terms of mutation rates and selection coefficients. The main conclusion, based on the estimates of possible parameter values, is that genetic instability is an early event with likely causal significance in the process of somatic mutation leading to cancer.

## Conclusions

Mathematical modeling always had been trying to find a compromise between simplicity of analysis and requirements of realism. On the one hand, we have extremely complex biological systems; on the other hand, we need to formally address some quantitative issues about these systems which often can be done only through the use of mathematical models that may rest on grossly over-simplified assumptions. On some occasions, however, a particular mathematical formalism seems to be “pre-adapted” to a variety of biological systems and can be profitably used to model a diverse set of processes. Birth-and-death stochastic processes are one class of such “lucky” models, as we attempted to show in this review using examples from several disparate areas of biology.

The utility of the theory of birth-and-death processes seems to be owing to its two crucial features. First, a huge variety of fundamental biological processes can be described in terms of elementary events that can be identified with birth and death event. These include actual birth of a new individual, cell division, mutation, gene duplication, horizontal gene transfer from one individual of a population to another (or from one population to another), appearance of a new lineage (e.g., a species) in a genealogy, and others. Second, the mathematical theory is well studied, relatively simple, many results are known, and the technique is flexible such that it can be adjusted in cases when the standard birth-and-death process is insufficient to describe the analyzed process. The BDIM discussed here are one such case where the additional elementary process of innovation had to be introduced to adequately incorporate for the existing basic understanding of genome evolution [70]. Another, conceptually similar example comes from population biology. Populations can suffer dramatic declines from disease or food shortage but, perhaps surprisingly, such populations can survive for long periods of time and, although they may eventually become extinct, they can exhibit an apparently stationary regime. This behavior has been successfully modeled using the so-called birth-death-and-catastrophe process [12,93,94]. Another extension of a birth-and-death process is the situation when transitions from a state are allowed not only to neighboring states but also to some other states within the given limits. Often, such systems are governed by a quasi-birth-and-death process (QBD process). These processes extend the classical birth-and-death processes to the vectorial case, and the tri-diagonal generator of parameters is substituted by a block-tridiagonal matrix ([95] and references therein). Furthermore, diffusion approximations of birth-and-death processes have been developed and characterized simultaneously with the standard version ([5]; see [11] for a mathematical description). Very recently, a diffusion approximation of the BDIM has been developed as a specific application to the analysis of evolution of gene families [96,97].

It seems that, with the rapid accumulation of genomic data, which include not only sequences but also genome-wide information on gene expression, genetic interactions between genes, protein-protein interactions, regulatory networks, and more, the role of mathematical modeling in the new integrative (systems) biology is going to be indispensable and increasing for the foreseeable future. We believe that birth-and-death processes comprise an essential part of the mathematical framework of this new biology.

## Figure legends

Figure 1. Representation of a birth-and-death process as a movement of a material particle between states. $\lambda_n \Delta t$ is the probability of a jump to the right from state $n$ (birth) and $\mu_n \Delta t$ is the probability of a jump to the left (death) during a short time interval $\Delta t$.

Figure 2. The fit of the observed size distribution of paralogous gene (domain) families in the human genome to the second-order balanced, linear BDIM (solid line). X axis: number of members in a family (family size); Y axis: number of families of a given size. The scale is double-logarithmic. The figure is from [70].

Figure 3. The time required for the formation of a first family with 1024 members (approximately, the size of the largest paralogous families in eukaryotic genomes) starting from an ensemble of 3000 singletons (blue), obtained by computer simulation, compared to the mean time predicted by BDIMs of different degrees (magenta). X axis: degree of the model; Y axis: time in billions of years estimated using the gene duplication rate from [42]. The figure is from [74].

Figure 4. (a) Contour plot of the average population penetration of a new (horizontally transferred) gene with the fixed value $uN = 0.5$. (b) Level lines for 50% population penetration of a new gene for different values of $uN$

$$0 \underset{\mu_1}{\overset{\lambda_0}{\rightleftarrows}} 1 \underset{\mu_2}{\overset{\lambda_1}{\rightleftarrows}} \cdots \underset{\mu_{n-1}}{\overset{\lambda_{n-2}}{\rightleftarrows}} n-1 \underset{\mu_n}{\overset{\lambda_{n-1}}{\rightleftarrows}} n \underset{\mu_{n+1}}{\overset{\lambda_n}{\rightleftarrows}} n+1 \underset{\mu_{n+2}}{\overset{\lambda_{n+1}}{\rightleftarrows}} \cdots \underset{\mu_N}{\overset{\lambda_{N-1}}{\rightleftarrows}} N$$

Novozhilov et al., Fig. 1

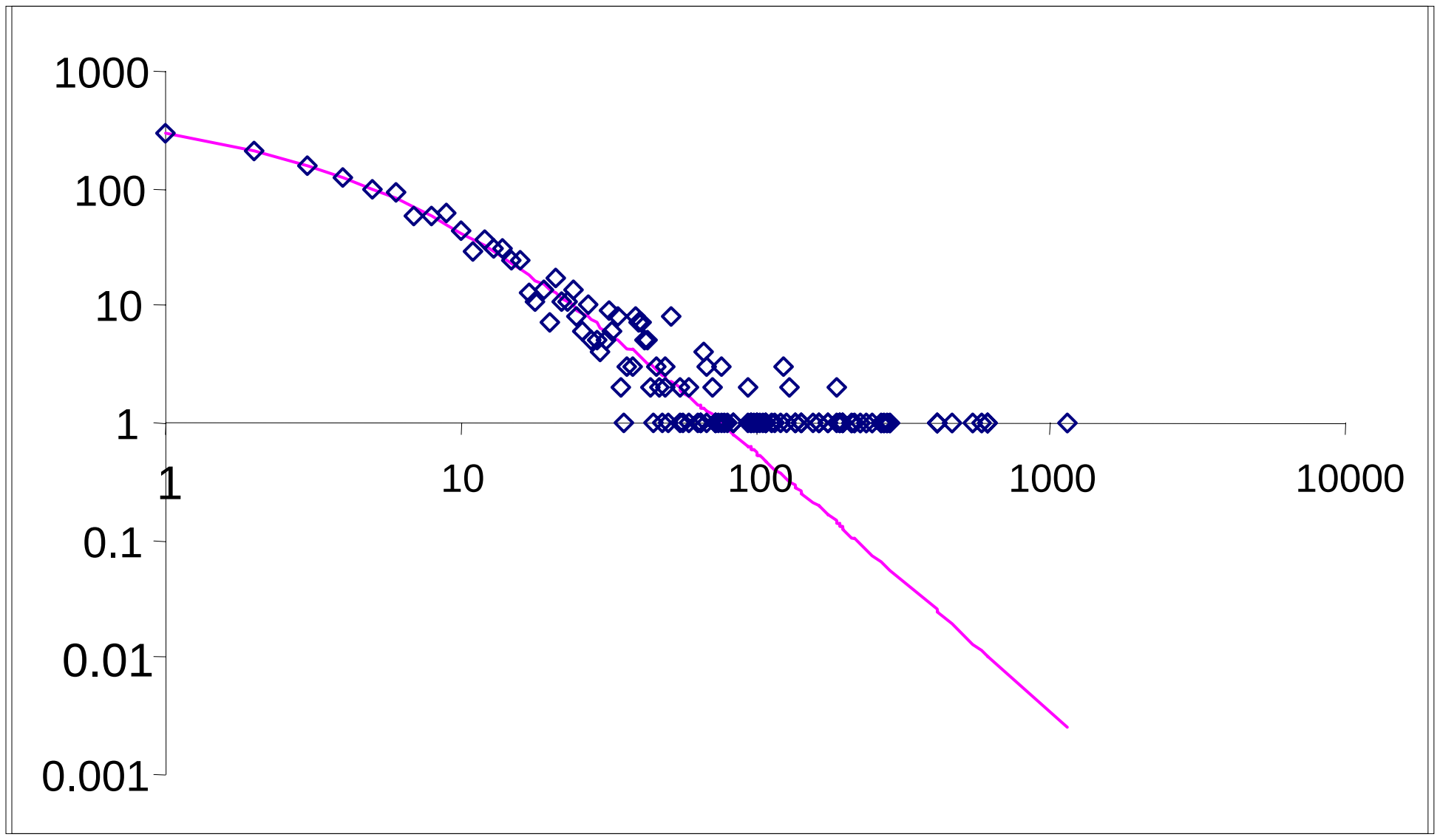

Novozhilov et al., Fig. 2

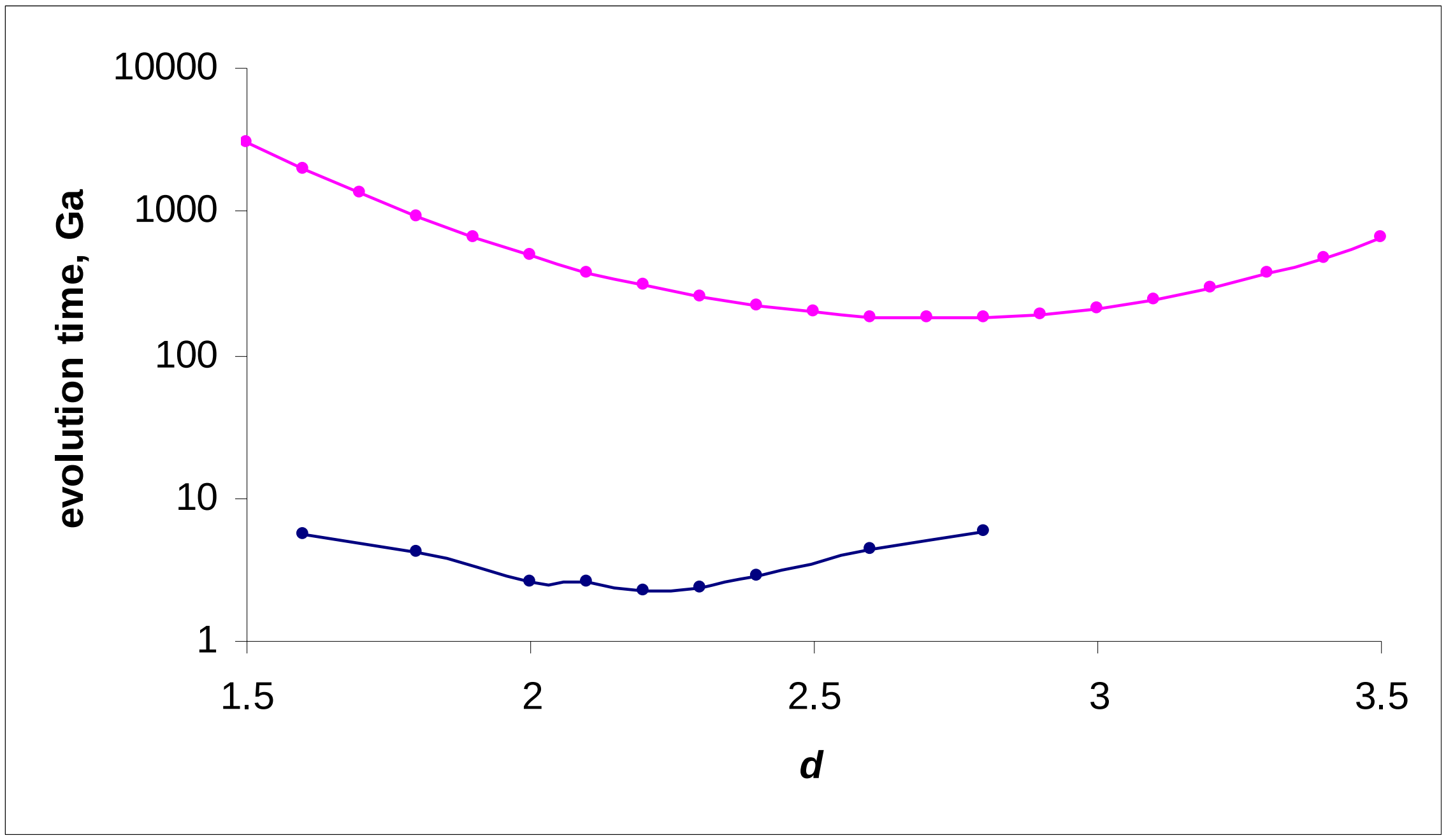

Novozhilov et al., Fig. 3

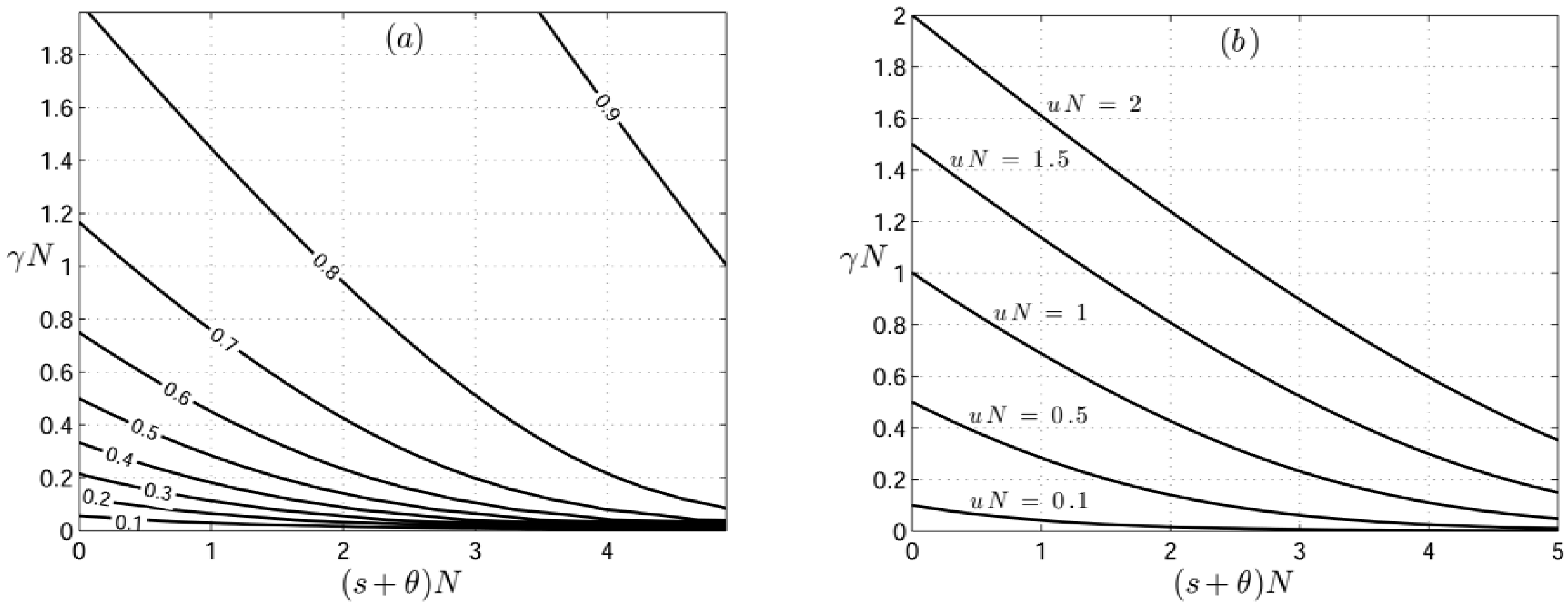

Novozhilov et al., Fig. 4